\documentstyle[aps,epsfig,floats,amssymb]{revtex}
\draft

\begin{document}
\twocolumn[\hsize\textwidth\columnwidth\hsize\csname
@twocolumnfalse\endcsname

\title{Can Structure Formation Influence the Cosmological Evolution?}

\author{Christof Wetterich}

\address{
Institut f{\"u}r Theoretische Physik,
Philosophenweg 16, 69120 Heidelberg, Germany}

\maketitle

\begin{abstract}
The backreaction of structure formation influences 
the cosmological evolution equation for the homogenous and isotropic average metric.
In a cold dark matter universe this effect leads only to
small corrections unless a substantial fraction of matter is located
in regions where strong gravitational fields evolve in time. A``cosmic virial
theorem'' states that the sum of gravitational and matter pressure vanishes
and therefore relates the average kinetic energy to a suitable average  of the
Newtonian potential. In presence of a scalar ``cosmon'' field mediating
quintessence, however, cosmology could
be modified if local cosmon fluctuations  grow large. We speculate that this
may trigger the accelerated expansion of the universe after the
formation of structure.
\end{abstract}
\pacs{PACS numbers: 98.80.Cq, 95.35.+d  \hfill HD-THEP-01-43}

 ]


\section{Introduction}

Recent observations of the Hubble diagram for supernovae 1a indicate that the
expansion of the universe may be accelerating in the present epoch \cite{SN}.
In this case the previous decrease of the Hubble parameter $H\sim t^{-1}$
must have slowed down just in the last couple of billion
years and an obvious question
asks: Why does this slowdown happen\footnote{On a logarithmic scale
as relevant for cosmology the last few billion years are more
or less the ``present'' epoch.} 
``just now''? A possible explanation
would be a cosmological constant which sets a mass scale $\lambda\approx
(10^{-3}{\rm eV})^4$ and therefore also a corresponding time scale 
$\sim (\lambda/M_p^2)^{-1/2}$, with $M_p\approx 10^{19}$ GeV
the Planck mass. Since it seems to be very hard to understand the 
origin of the tiny mass scale $\lambda$ theoretically, one is tempted
to look for alternatives. A possible scenario are models of quintessence
\cite{CW1},\cite{Q}. They are based on the time evolution of a
scalar field -- the cosmon -- which  is of cosmological relevance today.
In the simplest viable models, however, the characteristic
time for the onset of
acceleration is put in by hand in the form of the effective scalar potential
or kinetic term\footnote{For more recent examples see \cite{AS}, \cite{DMQ},
\cite{HE}.}. This does not always need tremendous fine-tuning of the order
of 100 digits as for the case of the cosmological constant. Indeed,
there are models where it is sufficient to tune parameters on the level
of percent to permille. We feel, nevertheless, that these ideas would become
much more credible if a natural solution of the ``why now'' problem
could be given. As one possibility the relevant time scale may be linked to
a natural small number arising from a fundamental theory. A recent proposal in
this direction involves the properties of a conformal fixed point \cite{CWFP}. As an
alternative, some event in the more
recent cosmological evolution could have ``set the clock'' to trigger
the acceleration at present. An idea in this direction \cite{MS}
-- ``$k$-essence'' -- tries to use the transition from a radiation-dominated
to a matter-dominated universe in order to set the clock.
Here we explore if structure formation could have induced a change of the pace of expansion.

One of the most striking qualitative changes in the recent history
of the universe is the formation of structure. For most of the cosmological
evolution the universe was homogenous to a high degree. Looking at the sky
today we see, however, strong inhomogeneities in the form of stars,
galaxies and clusters on length scales sufficiently small as compared
to the horizon. Could the emergence of the inhomogeneities set the
clock \cite{CWR} for the present acceleration?
In order to answer this question, we have to understand how
inhomogeneities on the scales of clusters or smaller ``act back'' on the
evolution of the homogenous ``average metric''. (For the purpose of
this paper we consider formally an average over the present horizon.
More accurately, the supernovae results concern an average over a
volume corresponding to $z\approx 1$.) After all, the universe is
not homogenous at present and the Einstein equations determine
the metric in presence of these inhomogeneities. One can still
formulate a type of ``macroscopic Einstein equation'' for the average
metric which, by definition, can be considered as homogenous. The
macroscopic equation simply obtains by averaging the ``microscopic
Einstein equation''. In this averaging procedure the ``backreaction''
of the inhomogeneities appears in the form
of ``correction terms'' in the macroscopic Einstein equation \cite{BR}. In
models of quintessence this holds also for the macroscopic evolution
equation for the scalar field.

It is the aim of the present paper to estimate the size and therefore
the relevance of the backreaction effects. For this purpose we express
in sect. 2 the backreaction effects in terms of a ``gravitational
energy density'' $\rho_g$ and corresponding ``pressure'' $p_g$. It is obvious that
$\rho_g$ and $p_g$ are relevant only if they are not tiny as compated
to the energy density $\bar\rho$ in radiation or matter. A very rough
estimate shows that in early cosmology the effects of $\rho_g$
and $p_g$ are indeed completely negligible. One the other hand,
once stars and galaxies have formed the ratio $\rho_g/\bar\rho$ is not
many orders below one any more, and a more detailed investigation
becomes necessary. In sect. 3 we evaluate $\rho_g$ and $p_g$
in terms of the correlation function for the local energy momentum
tensor of matter and radiation. This form exhibits clearly the
relation of these quantities to the inhomogeneities.

In sect. 4 we attempt a quantitative estimate for a standard
cold dark matter universe (without quintessence, but possibly
in presence of a cosmological constant). We find that the
effects of inhomogeneities on the scales of stars and galaxies
are small, they contribute typically $\rho_g/\bar\rho\approx 10^{-6}$.
A typical contribution from inhomogeneities on the scales of clusters
is $\rho_g/\bar\rho\approx 10^{-4}$. These estimates hold, however,
only if the fraction of matter in regions of strong gravitational
fields like black holes or the center of galaxies is small.
In sect. 5 we address the backreaction effects from black holes
and similar objects. The gravitational energy density $\rho_g$ can indeed
be large. Nevertheless, the combined energy momentum tensor for gravitational
and matter contributions behaves as for a nonrelativistic gas
if the objects are static. We conclude that for a cold dark matter
universe the backreaction effect could play a significative role
only if a substantial fraction of matter is found in regions where
strong gravitational fields evolve in time. This does not seem
to be very likely.

In models with quintessence the situation could change dramatically,
but only if the inhomogeneities in the cosmon field are substantial.
The ``gravitational backreaction'' $\rho_g, p_g$ is then supplemented
by a ``cosmon backreaction'' $\rho_c,p_c$
due to the cosmon fluctuations. We discuss a simple collection of
static and isotropic cosmon lumps in sect. 6. This would behave
similar to black holes. We argue that for more general, in particular
non-static, cosmon fluctuations the fine cancellation between
$p_c$ and $p_g$ observed in the static isotropic solutions may not be
maintained. In particular, it seems conceivable for a ``cosmon
dark matter'' scenario \cite{CDM} that $\rho_c/\rho$ and for
$\rho_g/\rho$ are of order unity. Under this condition
it would become quite likely that the formation of structure would lead
to a qualitative change in the evolution equation for the average metric.
One would expect deviations from $H\sim t^{-1}$ once structure has
formed. We summarize our conclusions in sect. 7.

\section{The influence of structure\protect\\ on the cosmological equations}

After the formation of structure the universe does not remain
homogenous on small scales. Nevertheless, we believe that
homogeneity and isotropy are realized on large scales and
describe the cosmological evolution by a Robertson-Walker metric.
The true metric $g_{\mu\nu}$ of the universe has to reflect the
inhomogeneities due to stars, galaxies and clusters. Therefore
the homogenous cosmological metric can at best be interpreted
as some type of average metric\footnote{The averaging is done here
with respect to the background metric. See ref. \cite{BU} for a recent discussion
of averaging procedures in a more general context.}
 $\bar g_{\mu\nu}=<g_{\mu\nu}>$. This
situation introduces a mismatch in the standard treatment
of the cosmological Einstein equations. On the right-hand side
one uses the average of the energy momentum $<t_{\mu\nu}>$,
whereas for the left-hand side one employs the Einstein tensor
formed from the average metric $\bar g_{\mu\nu}$. The correct
averaged Einstein equation involves\footnote{We use signature
(-,+,+,+), $R_{\mu\nu\rho}^{\ \ \ \ \lambda}=-\partial_\mu\Gamma_{\nu
\rho}^{\ \ \lambda}+...$ and $M^2=M^2_p/(16\pi)=1/(16\pi G_N)$.},
however, the averaged value of the Einstein tensor
\begin{equation}\label{2.1}
<R_{\mu\nu}-\frac{1}{2} Rg_{\mu\nu}>=\frac{1}{2M^2}<t_{\mu\nu}>\end{equation}
The difference between the averaged Einstein tensor and the Einstein
tensor formed from the average metric, i.e. $\bar R_{\mu\nu}-\frac{1}{2}
\bar R\bar g_{\mu\nu}$, introduces a correction term in the cosmological
equation for the average metric
\begin{equation}\label{2.2}
\bar R_{\mu\nu}-\frac{1}{2}\bar R\bar g_{\mu\nu}=\frac{1}{2M^2}T_{\mu\nu}
=\frac{1}{2M^2}(<t_{\mu\nu}>+T^g_{\mu\nu})\end{equation}
Here the ``gravitational correction'' to the energy momentum tensor
\begin{equation}\label{2.3}
T^g_{\mu\nu}=-2M^2<\delta G_{\mu\nu}>\end{equation}
\begin{equation}\label{2.4}
\delta G_{\mu\nu}=R_{\mu\nu}-\frac{1}{2} R g_{\mu\nu}-
(\bar R_{\mu\nu}-\frac{1}{2}\bar R\bar g_{\mu\nu})\end{equation}
reflects the influence of the inhomogeneities. It accounts for the
``backreaction'' of structure formation on the evolution
of the homogenous ``background metric'' $\bar g_{\mu\nu}$.
Homogeneity and isotropy of all averaged quantities imply that the only
nonvanishing components of $T^g_{\mu\nu}$ are given by
\begin{eqnarray}\label{2.5}
T_{00}^g&=&\rho_g=-2M^2<\delta G_{00}>\nonumber\\
T_{ij}^g&=&p_g\bar g_{ij}=-2M^2<\delta G_{ij}>
\end{eqnarray}
The cosmological equation therefore preserves its form, but
$T_{00}$ is not given solely by the average of the energy density
in matter and radiation. It also contains a gravitational contribution
which reflects the imprint of structure formation on inhomogeneities
of the metric. We observe that for a flat background metric
$\bar g_{\mu\nu}=\eta_{\mu\nu}$ the quantity $T_{\mu\nu}^{\ \ g}$
represents precisely the definition of the gravitational energy momentum
densities \cite{Wein}. Our setting is therefore a straightforward
generalization to cosmology.

At this point some comments about our averaging procedure seem in order. Assume that
in a given suitable gauge - we will later specify a particular one - the
detailed inhomogeneous geometry of the universe is described by the microscopic metric
$g_{\mu\nu}(\vec{x},t)$. It is related to the microscopic energy momentum tensor
$t_{\mu\nu}(\vec{x},t)$ by the microscopic Einstein equation.
\begin{equation}\label{2.5A}
R_{\mu\nu}-\frac{1}{2}Rg_{\mu\nu}=\frac{1}{2M^2}
t_{\mu\nu}\end{equation}
We next take a reference metric $\tilde{g}_{\mu\nu}$ of the homogeneous and isotropic
Robertson-Walker form $ds^2=-dt^2+a^2(t)d\vec{x}^2$. The microscopic metric can be
written as $g_{\mu\nu}=\tilde{g}_{\mu\nu}+h_{\mu\nu}$. Our reference metric defines
surfaces of fixed $t$. At any time $t$ we define the average by
\begin{equation}
\langle A\rangle(\vec{x},t)=\frac{1}{V}\int_Vd^3yA(\vec{y}-\vec{x},t)
\end{equation}
with $V$ a very large volume (typically of horizon size) in the comoving coordinates
\footnote{Since we average at fixed $t$ it does actually not matter if we average over
coordinates $a(t)\vec{x}$ or $\vec{x}$ if the physical averaging volume grows
$\sim a^3$.}. We can then fix the scale factor $a(t)$ in $\tilde{g}_{\mu\nu}$
self-consistently by requiring \footnote{In order to achieve this task we may use the
freedom of selecting a suitable gauge. Of course, $\langle h_{\mu\nu}\rangle=0$ is only
possible if the universe is really homogeneous and isotropic in average. In particular,
$\langle h_{\mu\nu}\rangle$ should not depend on $\vec{x}$ (after using the gauge freedom).}
$\langle h_{\mu\nu}\rangle=0$. This finally identifies $\tilde{g}_{\mu\nu}$ and
$\langle g_{\mu\nu}\rangle$.

Spatial averaging at fixed $t$ has been proposed by Futamase \cite{Fut1}. We are free
to use such an averaged description - the only physical assumption in our paper concerns
averaged homogeneity and isotropy of the real universe, namely in our paper
$\langle h_{\mu\nu}\rangle =0$ and $\langle t_{\mu\nu}\rangle$ depending only on $t$. A much
more subtle point is the question to what extent a real observer actually observes the
``spatially averaged'' quantities in the way introduced here. Detailed studies conclude
\cite{Fut2} that this may actually be the case - we will not address this topic in
the present paper. For a particular picture of the cold dark matter scenario Futamase
concludes that backreaction effects are small whereas Buchert et al. speculate
\cite{BU} that the influence on the cosmological evolution could be substantial,
nevertheless.

Let us next discuss the general structure of $\rho_g$ and $p_g$ (eq.
\ref{2.5}). As long as gravity remains weak, one can expand in the
small inhomogeneities of the metric
\begin{equation}\label{2.6}
g_{\mu\nu}=\bar g_{\mu\nu}+h_{\mu\nu}\end{equation}
such that
\begin{equation}\label{F2}
\delta G_{\mu\nu}=D^{\alpha\beta}_{\mu\nu}h_{\alpha\beta}
+E^{\alpha\beta\gamma\delta}_{\mu\nu}h_{\alpha\beta}
h_{\gamma\delta}\end{equation}
Here the differential operators $D$ and $E$ involve two
derivatives acting on $h$ or $\bar g$. They will be
computed more explicitly in sect. 3. From $<h_{\alpha\beta}>\equiv0$ one
concludes that $\rho_g$ and $p_g$ are quadratic in $h$,
\begin{eqnarray}\label{F3}
\rho_g&=&-2M^2<E^{\alpha\beta\gamma\delta}_{00}h_{\alpha\beta}
h_{\gamma\delta}>\nonumber\\
p_g&=&-\frac{2M^2}{3a^2}<E^{\alpha\beta\gamma\delta}_{ii}h_{\alpha\beta}
h_{\gamma\delta}>\end{eqnarray}
Thus $\rho_g$ and $p_g$ involve the correlation function for the metric and
do, in general, not vanish.

The local variation of the metric
reflects the local variations of the energy momentum
tensor according to the
``microscopic'' Einstein equation (\ref{2.5A}) \footnote{``Microscopic''
means here the scales of stars or galaxies...! These scales
are to be compared with the
``macroscopic'' scale of the order of the horizon.}
Within the linear approximation to eq. (\ref{2.5A}), namely
\begin{equation}\label{F6}
D^{\alpha\beta}_{\mu\nu}h_{\alpha\beta}=\frac{1}{2M^2}(t_{\mu\nu}
-\bar T_{\mu\nu})=\frac{1}{2M^2}\delta t_{\mu\nu}\end{equation}
the metric fluctuations $h_{\mu\nu}$ are linear in the
fluctuations of the energy momentum tensor $\delta t_{\mu\nu}$.
In consequence, $\rho_g$ can also be viewed as the effect of a nonvanishing
correlation function for the fluctuations of the energy density.
This correlation function can be observed as a galaxy -- or cluster -- correlation
function on appropriate length scales. In particular, we know that
on small scales the universe is far from homogenous. As an example,
on the length scale of the size of stars very dense regions (stars)
contrast with an almost empty environment. This is equivalent to
a huge correlation function and brings us back to the question:
Can the formation of stars or galaxies influence the evolution
of the universe as a whole?

In order to get a first rough estimate of the
magnitude\footnote{Note that this estimate does not account
for the total backreaction. Since we want to study here
the effects of structure formation we can concentrate on
a typical wavelength well within the horizon. Discussions
of the backreaction from modes outside the horizon can
be found in \cite{LSF}, \cite{CDM}.} of this
``backreaction of structure formation'' let us assume for a moment
that the universe consists of randomly distributed\footnote{Stars may
be replaced by galaxies or other extended objects.} ``stars'' with
radius $L$ or volume $v_L=\frac{4\pi}{3}L^3$ and density $\rho_L$.
Consider our horizon volume $V$ with $N_V$ stars. Since $T_{00}
\equiv \bar\rho$ is of the same order as $<t_{00}>=N_Vv_L\rho_L/V$
and, on the other side, $<t_{00}^2>=N_Vv_L\rho_L^2/V$, one has
\begin{equation}\label{F7}
\frac{<\delta\rho^2>}{\bar\rho^2}=\frac{<(t_{00}-\bar\rho)^2>}{\bar\rho^2}
\approx\frac{1}{f}\ ,\quad f=\frac{N_V v_L}{V}\approx\frac{\bar\rho}{
\rho_L}\end{equation}
The fraction of volume occupied by stars, $f$, is indeed a tiny number and
one concludes that the relative density fluctuations are huge! On the
other hand, the weak gravitational coupling enters such that the relative
size of $\rho_g$ as compared to $\bar\rho$ could still be small.
Since the operator $D$ in eq. (\ref{F6}) contains two derivatives a rough
estimate assumes
\begin{equation}
<h^2>\approx \frac{L^4}{M^4}<\delta\rho^2>\end{equation}
and, with a similar dimension argument,
\begin{equation}\label{F9}
\rho_g\approx\frac{L^2}{M^2}<\delta\rho^2>\approx\frac{L^2}{M^2}\rho_L\bar\rho
\end{equation}
The critical quantity for the relevance of the backreaction is therefore
given by the ratio
\begin{equation}\label{F10}
R=\frac{\rho_g}{\bar\rho}\approx\frac{L^2}{M^2}\rho_L
\approx\frac{m_L^{2/3}\rho_L^{1/3}}{M^2}
\approx\frac{m_L}{LM^2}\end{equation}
with $m_L$ the mass of the stars.
It is suppressed by two powers of the Planck mass as expected for
a gravitational fluctuation effect. On the other hand, the
mass $m_L$ of the star and its size $L$ are huge in
microphysical units. Inserting values typical for the sun, $m_L=2\cdot 10^{33}g
=1.1\cdot 10^{57}$ GeV, $L=7\cdot 10^8m=3.5\cdot
10^{24}$ GeV$^{-1}$ and using $M=1.72\cdot10^{18}$ GeV one finds for
main sequence stars
\begin{equation}\label{F11}
\frac{m_L}{LM^2}\approx 10^{-4}\end{equation}

Another estimate relates the gravitational backreaction effect to
typical values of the Newtonian gravitational potential $\phi =-h_{00}/2$
in extended objects. Indeed, we note that $R$ is proportional
to the gravitational potential at the surface of the star, $m_LG/L$,
with $G^{-1}=16\pi M^2$. Its value for the sun is
\begin{equation}\label{label}
-\phi=\frac{m_LG}{L}=2.12\cdot 10^{-6}
\end{equation}
Similarly, for idealized neutron stars with mass at the
Oppenheimer-Volkoff limit,
$m_L=1.4\cdot 10^{33}g,\ L=9.6 km$ one has
\begin{equation}\label{F11B}
\frac{m_L}{LM^2}=5.5\ ,\quad -\phi=\frac{Gm_L}{L}=0.11\end{equation}
These first estimates are, perhaps surprisingly, not much below
one (as could have been expected from the factor $M^{-2}$).
A more detailed investigation, including the various
proportionality constants and the distribution of objects with
different $L$ and $M_L$, becomes necessary.

Before doing so, it is instructive to discuss a few qualitative
aspects of the dependence of the ratio $R$ on $L$ and $\rho_L$:
(1) The ratio $\rho_g/\bar\rho$ is independent of $\bar\rho$.
It therefore shows essentially no time-dependence once the objects
have condensed with a stationary density and size. (2) For
a fixed density $\rho_L$ the contribution from smaller objects
vanishes rapidly. For example, the condensation to dust particles
or planets
is many orders of magnitude too small to be relevant. (3) Microphysical
objects like nuclei play no role for $\rho_g$ (i.e. $R\approx 10^{-36}$
for a gas of nuclei). In early cosmology the contribution of $\rho_g$ is
therefore completely negligible. (4) For an (elliptical) galaxy
consisting of $\nu_G$ roughly uniformly distributed stars 
within a radius $L_G$ the density scales $\rho_G\approx \nu_G(L/L_G)^3
\rho_L$. (There may be some moderate
enhancement from dark matter.) For a uniform
mass distribution in a galaxy the combination $L^2_G\rho_G=\nu_G(L/L_G)L^2\rho_L$
is changed by a factor $\nu_GL/L_G$ as compared to stars.

\section{Gravitational energy momentum tensor in cosmology}

We next turn to a more quantitative discussion of eqs. (\ref{F2})
and (\ref{F3}) for the metric inhomogeneities. Since the relevant
length scale for the dominant fluctuations is much smaller than the
horizon, we can neglect derivatives acting on $\bar g_{\mu\nu}$ as
compared to those acting on $h_{\mu\nu}$. This yields the
microscopic field equation up to quadratic order in the
metric fluctuations
\begin{eqnarray}\label{3.1}
\delta G_{\mu\nu}&=&-\frac{1}{2}\left\{\partial^\rho\partial_\rho
h_{\mu\nu}+\partial_\mu\partial_\nu h^\rho_{\ \rho}-\partial_\mu
\partial_\rho h^\rho_{\ \nu}-\partial_\nu\partial_\rho h^\rho_{\ \mu}
\right.\nonumber\\
&&\left.-\partial^\rho\partial_\rho h^\alpha_{\ \alpha}\bar g_{\mu\nu}
+\partial_\alpha\partial_\rho h^{\alpha\rho}\bar g_{\mu\nu}
\right\}\nonumber\\
&&+\frac{1}{2}h^{\alpha\rho}\left\{\partial_\mu\partial_\nu h_{\alpha\rho}
+\partial_\alpha\partial_\rho
h_{\mu\nu}-\partial_\rho\partial_\mu h_{\alpha\nu}\right.\nonumber\\
&&\left.-\partial_\rho\partial_\nu h_{\alpha\mu}\right\}
+\frac{1}{2} h_{\mu\nu}\left\{\partial^\rho\partial_\rho h^\alpha_{\
\alpha}-\partial_\rho\partial_\alpha h^{\alpha\rho}\right\}\nonumber\\
&&-\frac{1}{2}\bar g_{\mu\nu}h^{\alpha\rho}\left\{
\partial^\beta\partial_\beta h_{\alpha\rho}
+\partial_\alpha\partial_\rho h^\beta_{\ \beta}-2\partial_\rho
\partial_\beta h^\beta_{\ \alpha}\right\}\nonumber\\
&&+\frac{1}{4}\left\{\partial_\mu h^{\alpha\rho}\partial_\nu h_{\alpha\rho}
+2\partial_\alpha h^{\alpha\rho}\partial_\rho h_{\mu
\nu}-\partial^\rho h^\alpha_{\ \alpha}\partial_\rho h_{\mu\nu}\right.
\nonumber\\
&&+2\partial^\alpha h^\rho_{\ \mu}\partial_\alpha h_{\rho\nu}-2
\partial^\alpha h_{\rho\mu}\partial^\rho h_{\alpha\nu}
-2\partial_\alpha h^{\alpha\rho}\partial_\mu h_{\nu\rho}\nonumber\\
&&\left.-2\partial_\alpha h^{\alpha\rho}\partial_\nu
h_{\mu\rho}
+\partial^\rho h^\alpha_{\ \alpha}\partial_\mu h_{\nu\rho}+
\partial^\rho h^\alpha_{\ \alpha}\partial_\nu h_{\mu\rho}\right\}
\nonumber\\
&&-\frac{1}{8}\bar g_{\mu\nu}\left\{ 3\partial^\alpha
h_{\rho\beta}\partial_\alpha h^{\rho\beta}+4\partial_\alpha h^{\alpha\rho}
\partial_\rho h^\beta_{\ \beta}\right.\nonumber\\
&&\left.-\partial^\alpha h^\rho_{\ \rho}\partial_\alpha
h^\beta_{\ \beta}-4\partial_\alpha h^{\alpha\beta}\partial_\rho
h^\rho_{\ \beta}-2\partial_\alpha h^{\rho\beta}\partial_\rho h^\alpha
_{\ \beta}\right\}\nonumber\\
&=&\frac{1}{2M^2}\delta t_{\mu\nu}\end{eqnarray}
Here the indices are raised and lowered with the homogenous background
metric $\bar g_{\mu\nu}$. The average $<\delta G_{\mu\nu}>$ concerns
only the part quadratic in $h_{\mu\nu}$,
since  $<h_{\mu\nu}>$ vanishes by definition. It is homogenous (it involves
a volume integral) and we can therefore perform integration by parts
for the space derivatives. On time scales of the order of the characteristic
length scales of the fluctuations $<\delta G_{\mu\nu}>$ is also
essentially static. This allows us to perform integration by parts
for the time derivatives as well, and we obtain
\begin{eqnarray}\label{3.2}
&&<\delta G_{\mu\nu}>=\frac{1}{4}<h^{\alpha\rho}\partial_\mu\partial_\nu
h_{\alpha\rho}+3h^\alpha_{\ \alpha}\partial^\rho\partial_\rho h_{\mu\nu}
\nonumber\\
&&\hspace{1,5cm}-2h^{\alpha\rho}\partial_\rho\partial_\alpha h_{\mu\nu}-2h^\rho_{\ \mu}
\partial^\alpha\partial_\alpha h_{\rho\nu}+2h_{\rho\mu}
\partial^\rho \partial_\alpha h^\alpha_{\ \nu}\nonumber\\
&&\hspace{1,5cm}-h^\alpha_{\ \alpha}\partial_\mu\partial_\rho h^\rho_{\ \nu}-h^\alpha
_{\ \alpha}\partial_\nu\partial_\rho h^\rho_{\ \mu}>\nonumber\\
&&\hspace{1,5cm}-\frac{1}{8}\bar g_{\mu\nu}<h^{\alpha\rho}
\partial^\beta\partial_\beta h_{\alpha\rho}
+h^\alpha_{\ \alpha}\partial^\rho\partial_\rho h^\beta_{\ \beta}\nonumber\\
&&\hspace{1,5cm}-2h^{\alpha\rho}\partial_\rho\partial_\beta h^\beta_{\ \alpha}>\end{eqnarray}
On the other hand, the linear part of the field equation relates the metric perturbations
to the perturbations in $t_{\mu\nu}$
\begin{eqnarray}\label{3.3}
&&\partial^2h_{\mu\nu}+\partial_\mu\partial_\nu h-\partial_\mu\partial_\rho
h^\rho_{\ \nu}-\partial_\nu\partial_\rho h^\rho_{\ \mu}\nonumber\\
&&=-\frac{1}{M^2}(\delta t_{\mu\nu}-\frac{1}{2}\delta t^\rho_{\ \rho}
\bar g_{\mu\nu}) \nonumber\\
&&\partial^2 h-\partial_\mu\partial_\nu h^{\mu\nu}=\frac{1}{2M^2}\delta t^
\rho_{\ \rho}\end{eqnarray}
where we use $\partial^2=\partial^\rho\partial_\rho$ and $h=h^\rho_{\ \rho}$.

Eqs. (\ref{3.2}) and (\ref{3.3}) simplify considerably in the harmonic gauge
which we adopt from now on
\begin{equation}\label{3.4}
\partial_\mu h^\mu_{\ \nu}=\frac{1}{2}\partial_\nu h\end{equation}
The linear field equation becomes
\begin{equation}\label{3.5}
\partial^2 h_{\mu\nu}=-\frac{1}{M^2}(\delta t_{\mu\nu}-\frac{1}{2}
\delta t^\rho_{\ \rho}\bar g_{\mu\nu})
=-\frac{1}{M^2}\delta s_{\mu\nu}\end{equation}
and the quadratic metric fluctuations read
\begin{eqnarray}\label{3.6}
&&<\delta G_{\mu\nu}>=\frac{1}{4}<h^{\alpha\rho}
\partial_\mu\partial_\nu h_{\alpha\rho}-\frac{1}{2}
h\partial_\mu\partial_\nu
h-2h^\rho_{\ \mu}\partial^2 h_{\rho\nu}\nonumber\\
&&\qquad+2h\partial^2h_{\mu\nu}>
-\frac{1}{8}\bar g_{\mu\nu}<h^{\alpha\rho}\partial^2 h_{\alpha\rho}
+\frac{1}{2}h\partial^2h>\end{eqnarray}
Neglecting graviational waves eq. (\ref{3.5}) has the retarded solution
\begin{equation}\label{3.7}
h_{\mu\nu}(\vec x,\tau)=\frac{a^2}{4\pi M^2}\int d^3\vec x'\frac{\delta s_{
\mu\nu}(\vec x',\tau-|\vec x-\vec x'|)}{|\vec x-\vec x'|}\end{equation}
where $\tau$ obeys $d\tau=dt/a$.
We recover Newton's law for the graviational potential $\phi=-h_{00}/2$
for static point sources.

For a distribution of star-like objects the time derivatives of $h_{\mu\nu}$
involve the peculiar comoving velocities of these objects. Since the peculiar
velocities are small as compared to the speed of light, we can
neglect the time derivatives in eq. (\ref{3.6}) as compared
to the space derivatives. Furthermore, by virtue of rotation
symmetry the expectation values involving only one derivative in a
given space direction vanish and one infers
\begin{eqnarray}\label{3.8}
\rho_g&=&-2M^2<\delta G_{00}>
=-2M^2<\frac{1}{8}h^{\alpha\rho}\Delta h_{\alpha\rho}\nonumber\\
&&+\frac{1}{16}h\Delta
h-\frac{1}{2}h^\rho_{\ 0}\Delta h_{\rho0}+\frac{1}{2}h\Delta h_{00}>
\nonumber\\
p_g&=&-\frac{2M^2}{3}<\delta G^i_i>
=2M^2<\frac{1}{24}h^{\alpha\rho}\Delta h_{\alpha\rho}\nonumber\\
&&+\frac{5}{48}h\Delta
h+\frac{1}{6}h^{\rho i}\Delta h_{\rho i}-\frac{1}{6}h\Delta h^i_i>\end{eqnarray}
with $\Delta=\bar g^{ij}\partial_i\partial_j$.  Sums over all
double indices are implied, with
latin indices running from 1 to 3.
To leading order we only need to
keep $\delta t_{00}$ such that $\delta s_{00}=\frac{1}{2}
\delta t_{00}=\frac{1}{2}\delta\rho,\delta s_{ij}=\frac{1}{2} \delta\rho \bar g_{ij}$ and $h_{ij}=h_{00}\bar g_{ij}$,
$h_{0i}=0,\ h=2h_{00}$. This results in
\begin{eqnarray}\label{3.9}
\rho_g&=&-\frac{9}{2}M^2<h_{00}\Delta h_{00}>\nonumber\\
p_g&=&\frac{1}{6}M^2<h_{00}\Delta h_{00}>\end{eqnarray}
and we infer the ``equation of state'' for the gravitational energy
momentum tensors of starlike objects
\begin{equation}\label{3.10}
p_g=-\frac{1}{27}\rho_g\end{equation}

Using (\ref{3.7}) we can also express $\rho_g$ in terms of the
correlation function for the energy density fluctuations $(V=a^3\int d^3x)$
\begin{equation}\label{3.11}
\rho_g=\frac{9a^5}{32\pi M^2 V}\int d^3 x\int d^3y\frac{1}{|\vec x-\vec y|}\delta\rho(\vec x)\delta\rho(\vec y)\end{equation}
(The ``retardation'' in eq. (\ref{3.7}) can be neglected since it
involves  again the peculiar velocities.) It is instructive to employ
a comoving Fourier basis
\begin{equation}\label{3.12}
\delta\rho(x)=\int\frac{d^3k}{(2\pi)^3}
e^{i\vec k\vec x}\delta\rho(\vec k)\end{equation}
where
\begin{equation}\label{3.13}
<\delta\rho(\vec k)\delta\rho(\vec k')>=G(k)(2\pi)^3\delta(\vec k
+\vec k')\end{equation}
The two-point density correlation function $G(k)$ depends only on
the invariant $k^2\equiv \vec k^2$. One finds
\begin{equation}\label{3.14}
\rho_g=\frac{9 a^2}{8 M^2}\int\frac{d^3k}{(2\pi)^3}k^{-2} G(k)\end{equation}
From $\rho(-k)=\rho^*(k)$ one infers that $G(k)$ is a real positive
quantity. This implies that $\rho_g$ is positive whereas $p_g$ is negative.
For small $k$ or long distances $G(k)$ decreases rapidly and the $k$-integral
is infrared finite. The interesting part comes from large $k$, where
condensed objects like stars contribute. On these scales it is
convenient to switch to physical momenta $\vec q=\vec k/a$
such that
\begin{equation}\label{3.15}
\rho_g=\frac{9}{8M^2}\int\frac{d^3q}{(2\pi)^3} q^{-2}\tilde G(q)\end{equation}
Here we employ $\delta\rho(x)=\int\frac{d^3q}{(2\pi)^3}e^{ia\vec q\vec x}
\delta\rho(q)$ and $<\delta\rho(\vec q)\delta\rho(\vec q')>
=\tilde G(q)(2\pi)^3\delta(\vec q-\vec q')$ is the correlation function as a
function of physical (not comoving) momenta. For a given static
$\tilde G(q)$
the gravitational incoherent energy density $\rho_g$ would not depend
on the scale factor. However, the condensed objects are diluted by
the cosmological expansion, and $G(q)\sim a^{-3}$ implies
$\rho_g\sim a^{-3}$, similar to the energy density in dark or
baryonic matter.
We conclude that $\rho_g$ is essentially a fixed fraction of the energy density of matter
$:R=\rho_g/\bar{\rho}>0$ is independent of time.

Together with the gravitational equation of state (\ref{3.10}) this can
actually be used for an
estimate of corrections to the equation of state of matter, $\bar{p}=w_m\bar{\rho}$.
Indeed, the matter and gravitational energy momentum tensor are not separately conserved.
Gravitational potentials lead to pecular velocities and therefore to nonzero $\bar{p}$.
In other words, the equation of state $w_m=0$ holds only for ``free particles''
(ideal dust), i. e. if the gravitational interactions are neglected. As for all
interacting systems we expect corrections. Conservation of the total energy momentum
tensor is, of course, exact and implies for $dR/dt=0$
\begin{equation}\label{3.17}
\dot{\bar{\rho}}(1+R)+3H(\bar{\rho}+\bar{p}+\rho_g+p_g)=0
\end{equation}
If $\bar{\rho}$ is dominated by massive objects or massive nonrelativistic particles we
can approximate $\bar{\rho}=\bar{\rho}_M+3\bar{p}/2$ where $\bar{\rho}_M$ is the
contribution of the particle masses. Assuming that no masses are added or changed during
the relevant period in the cosmological evolution we infer $\bar{\rho}_M\sim a^{-3},
\dot{{\bar{\rho}}}_M=-3H\bar{\rho}_M$. Furthermore, if $\bar{p}/\bar{\rho}$ is
approximately constant this extends to $\dot{{\bar{\rho}}}=-3H\bar{\rho}$.
Eq. (\ref{3.17}) therefore yields the simple relation
\begin{equation}\label{35}
\bar{p}+\bar{p}_g=0
\end{equation}
and we infer an estimate for the pressure of matter which is due to the gravitational
interactions
\begin{equation}
\bar{p}=w_m\bar{\rho}=-\frac{p_g}{\rho_g}R\bar{\rho}
=\frac{R}{27}\bar{\rho}
\end{equation}
Since $R$ is small (see the next section for an estimate) this amounts only to a tiny
correction, justifying our neglection of pecular velocities. We note that the estimate
(\ref{35})(\ref{3.10}) plays the role of a ``cosmic virial theorem'' since it
relates the average
kinetic energy $(\bar{p})$\footnote{The value of $\rho_g$ depends on the precise
definition of this
quantity, e. g. $\tilde{\rho}_g=2M^2\langle G_{0\rho}g^{\rho 0}\rangle =\rho_g-2M^2
\langle G^{(1)}_{0\rho}h^{\rho 0}\rangle=(5/9)\rho_g$. This does not affect $p_g$ and the
relation $\bar{p}+p_g=0$ whereas the ratios $p_g/\rho_g$ and $\rho_g/\bar{\rho}$ get modified.}
to the average gravitational potential $(\rho_g)$.

Turning our argument around we emphasize that the relation (\ref{35}) implies the
``cold dark matter expansion law'' $\bar{\rho}\sim a^{-3}$ provided $\dot{R}=0$.
If we would neglect $p_g$, a nonzero pressure $\bar{p}$ would correct the expansion
according to $\bar{\rho}\sim a^{-3(1+w_m)}$. This correction is cancelled by the
presence of the ``gravitational pressure'' $p_g$. Thus backreaction effects play a role
for the evolution - its role being to ensure the validity of the cold dark  matter expansion
law even in presence of pecular velocities or nonzero $\bar{p}$. Corrections arise only for
periods where $\dot{R}$ or $\dot{w}_m$ do not vanish.

\section{Do stars and galaxies modify the expansion of the universe?}

In this section we estimate the size of the backreaction effect
quantitatively for a standard cold dark matter universe.
We can use eq. (\ref{3.5})  in order to express $\Delta h_{00}$ in
terms of $\delta\rho$ and obtain from eq. (\ref{3.9})
\begin{equation}\label{4.1}
\rho_g=\frac{9}{4}<h_{00}\delta\rho>\end{equation}
For small $h_{00}$ this ``weighs'' the energy contrast $\delta\rho$
with the Newtonian potential $\phi$
\begin{equation}\label{4.2}
\rho_g=-\frac{9}{2}<\phi\delta\rho>\end{equation}
For compact objects $\delta\rho$ is almost equal to
the local value of $\rho$. For starlike extended objects the
size of their own gravitational potential is maximal at the surface,
$\phi_{max}=-mG/L$. For small $\phi_{max}$ the contribution of isolated
stars to $\rho_g$ is therefore suppressed
by a factor $\sim 10^{-6}$ as compared to their contribution
to $\bar\rho$, in accordance with eqs.  (\ref{F10}), (\ref{label}), (\ref
{F11B}). We need, however, also the contribution of other stars to
$\phi$. This becomes particularly simple in the language (\ref{4.1})
or (\ref{4.2}). As long as gravity remains weak, we only have to
fold any mass concentration with the gravitational potential at
the same location. Incidentally, this shows that our previous association
of the relevant ratio $R=\rho_g/\bar\rho$ with the Newtonian potential
can be made quantitative
\begin{equation}\label{4.3}
R=-\frac{9}{2}\ll\phi\gg=-\frac{9\langle\phi\delta\rho\rangle}{2\bar{\rho}}
\end{equation}
where $\ll\phi\gg$ means an appropriately weighted value of $\phi$.
This also yields a quantitative value for the pressure of matter (and therefore the
kinetic energy or pecular velocities) according to the ``cosmic viral theorem''
\begin{equation}\label{4.3A}
\bar{p}=-\frac{1}{6}\ll\phi\gg
\end{equation}
Note that equilibration has not been invoked for this estimate. The cosmic virial theorem
follows directly from $p_g/\bar{\rho}=\ll\phi\gg/6, \dot{R}=0$ and
$\bar{p}/\bar{\rho}=const.$. The average kinetic energy density
$\langle T\rangle /V=3\bar{p}/2=-\ll \phi \gg /4$ may be compared
with a virialized gravitationally bound system where $\langle T\rangle /V=-
\langle\phi\rangle /2$.

For cold dark matter galaxies the value of the galactic gravitational potential
in the outer regions, in particular the halo, can be estimated from the
rotation velocities
\begin{equation}
v_{rot}^2(r)=r\frac{\partial}{\partial r}\phi
\end{equation}
Within the halo $(r\leq r_H)$ the dependence of $\phi$ on $r$ is
approximately logarithmic
\begin{equation}
\phi=-\bar v_{rot}^2\ln\frac{r_H}{r}
\end{equation}
With $v_{rot}=0(10^{-3})$ we conclude that the galactic potential is
of similar size as the local potential on the surface of a typical star (\ref{label}).

Clusters of galaxies, however, have a deeper potential well. A typical value
for a cluster is
\begin{equation}
\phi_{cl}=-10^{-4}
\end{equation}
If most matter is found within clusters, this gives an approximate lower
bound for the gravitational energy density
\begin{equation}\label{4.7}
\rho_g\stackrel{\scriptstyle>}{\sim}\frac{9}{2}
|\phi_{cl}|\bar\rho\end{equation}
We observe that this effect results from the mutual coherent correlations
between all the stars in a cluster. The dominant length scale of
this contribution to the correlation function (\ref{3.15}) is related
to the size of the cluster.

There may still be sizeable contributions arising
from correlations on smaller
scales. The center of the galaxy typically contains a region with large
gravitational field. In this region, however, our linearized analysis
does not apply any more. A similar statement holds for individual
black holes outside the center of the galaxy. The precise evaluation
of these contributions to $\rho_g$ needs a nonlinear analysis and
depends crucially on the question how much of the matter in the universe
is found in regions with a strong gravitational field.
We will briefly turn to this question in the next
section. Only if such ``strong field contributions''
are substantially above the bound (\ref{4.7}),
the gravitational
energy density could be relevant for the evolution
of the universe. On the other hand, for a moderate contribution
from strong field regions the backreaction effect remains
small for conventional dark matter galaxies and clusters. A value
$\rho_g/\bar\rho\stackrel{\scriptstyle <}{\sim} 10^{-3}$
seems to be too small to
substantially modify the evolution of the universe
after structure formation.

\section{Contribution of black holes}

For black holes and other regions with strong gravitational
fields the linear analysis of the preceeding sections does not remain
valid. For an individual black hole -- or any other static and isotropic
object -- in a flat space-time background the sum of matter and
gravitational energy density  is fixed, however, by a conservation
law. This also holds for the pressure.
These laws can be expressed in terms of linearized gravity \cite{Wein}
and are the analogue of charge conservation in electromagnetism.
We parametrize the static and isotropic metric outside
a mass concentration in ``isotropic coordinates'' as $ds^2=-B(u)dt^2+C(u)d
\vec x d\vec x$ with $u^2=\vec x\vec x$. The sum of the energy densities
and the total pressure are related to the functions $B$ and $C$ by the
linearized Einstein equations, with $C'=\partial C/\partial u$, etc., as
\begin{eqnarray}\label{N5.1}
\rho(u)+\rho_g(u)&=&-2M^2(C''+\frac{2}{u}C')\nonumber\\
p(u)+p_g(u)&=&-\frac{2M^2}{3}(C''+\frac{2}{u}C'+B''+\frac{2}{u}B')\end{eqnarray}
Also, using Gauss' law, one finds for the integrals over a volume with
$u'<u$:
\begin{eqnarray}\label{N5.2}
m(u)&=&4\pi\int du'{u'}^2(\rho(u')+\rho_g(u'))=-8\pi M^2u^2C'(u)\nonumber\\
\hat P(u)&=&4\pi\int du'{u'}^2(p(u')+p_g(u'))\nonumber\\
&=&\frac{1}{3} m(u)
-\frac{8\pi}{3}M^2u^2B'(u)\end{eqnarray}
For the Schwarzschild metric the functions $B(u)$ and $C(u)$ are given
by $(G^{-1}=16\pi M^2)$
\begin{eqnarray}\label{N5.3}
B(u)&=&\left(1-\frac{mG}{2u}\right)^2\left(1+\frac{mG}{2u}\right)^{-2},
\nonumber\\
C(u)&=&\left(1+\frac{mG}{2u}\right)^4\end{eqnarray}
This yields, in particular, $m(u\to\infty)=m$ with $m$ the total
mass of the object related to the Schwarzschild radius $R_S=m/(8\pi M^2)$.
Similarly, we observe that the integrated pressure vanishes,
$\hat P(u\to\infty)=0$.

On a length scale which is large as compared to the characteristic
size of the objects a collection of static isotropic objects
-- including black holes -- can be viewed as a collection of point particles
with masses $m_l$. The total energy momentum tensor $T_{\mu\nu}$ in
eq. (\ref{2.2}) averages both the matter and gravitational contributions.
If the objects are sufficiently distant from each other, this amounts
to summing $m_l(u\to\infty)$ and $\hat P_l(u\to\infty)$. A collection
of static isotropic objects behaves therefore like a
nonrelativistic gas with zero pressure\footnote{The vanishing
of the pressure including the gravitational contribution $p_g$ is actually
even better obeyed as if $p_g$ had been neglected. Note that $<t_{\mu\nu}>
+T^g_{\mu\nu}$ is covariantly conserved with respect to the
background metric $\bar g_{\mu\nu}$ by virtue of eq. (\ref{2.2}).}.
In particular, black holes that have already formed before structure
formation -- this is the meaning of ``static'' in a
cosmological context -- behave just as a contribution to cold dark
matter. Irrespective of the fact that their gravitational
energy density $\rho_g$ can be substantial, the backreaction
effect from condensed black holes during or after structure formation
would not lead to a deviation from the usual equation of state.

The only loophole in the argument that backreaction effects can be
neglected in a cold dark matter universe remains the hypothesis of a substantial
contribution from black holes forming during or
after structure formation. For such objects we cannot use the static
approximation (\ref{N5.2}). At this stage we cannot exclude that a nonzero,
perhaps even negative, pressure could play a role for non-static
regions with strong gravitational fields.

For a cold dark matter universe we conclude that a sizeable influence of
backreaction effects is only possible if a substantial fraction of the energy
density is found in regions of strong gravitational fields which evolve
in time and cannot be described effectively as static objects. For being
important at the time relevant to the supernovae Hubble diagram such a
hypothetical evolution would have to persist at a redshift $z\approx 1$.
Discarding this -- perhaps rather unlikely -- possibility we find no
relevant backreaction effect from structure formation in a cold dark matter
universe.

\section{Cosmon fluctuations}

Recent cosmological observations suggest the presence of a homogenous dark
energy component. It has been proposed  that the dark energy
density is time-dependent and can be described by the dynamics of
a scalar field, the cosmon \cite{CW1}. If this ``quintessence'' scenario
\cite{CW1}, \cite{Q}
is true, one may also suspect that inhomogeneities in the cosmon field
could be associated with extended structures \cite{CL}. In this section
we argue that the backreaction effect of structure formation is much stronger
in a ``cosmon dark matter universe'' \cite{CDM} than in the
standard cold dark matter universe. One main reason
is the direct contribution of cosmon fluctuations
to the averaged energy momentum tensor. One also observes
large space-like
components of the gravitational field in cosmon lumps,
contributing to large $\rho_g$ and $p_g$. We underline
that the material of this section is only relevant if the present
local fluctuations of the cosmon field are really substantial
-- a possibility  that remains speculative as long as no consistent
picture of a cosmon dark matter universe has been developed. If the scalar
field mediating quintessence remains homogenous to a high degree
in the present epoch, its ``backreaction'' effects are small and can
be neglected.

There are three new ingredients for the backreaction in presence
of an inhomogenous scalar field:

(1) Local fluctuations of the scalar field
around its homogenous background value induce a new contribution to the
total energy momentum tensor (\ref{2.2}). The scalar contribution
to the local energy momentum tensor
\begin{equation}\label{N6.A}
t^\varphi_{\mu\nu}=-V(\varphi)g_{\mu\nu}+\partial_\mu\varphi
\partial_\nu\varphi-\frac{1}{2}\partial^\rho\varphi\partial_\rho\varphi
g_{\mu\nu}\end{equation}
yields, after averaging in eq. (\ref{2.1}) or (\ref{2.2}), both
a contribution from the homogenous background field $\bar\varphi$
(the average of $\varphi$) and from the inhomogenous local
fluctuations of the cosmon field $\delta\varphi=\varphi-\bar\varphi$,
namely
\begin{equation}\label{N6.B}
<t^\varphi_{\mu\nu}>=T^h_{\mu\nu}+T^c_{\mu\nu}\end{equation}
Here $T_{\mu\nu}^h$ stands for the time variable dark energy
or homogenous
quintessence and corresponds to eq. (\ref{N6.A}) with $\varphi$
replaced by $\bar\varphi$
\begin{eqnarray}\label{N6.C}
T^h_{00}&=&\rho_h=V(\bar\varphi)+\frac{1}{2}\dot{\bar\varphi}^2\,\nonumber\\
T^h_{ij}&=&p_h\bar g_{ij}\ ,\quad p_h=-V(\bar\varphi)+\frac{1}{2}
\dot{\bar\varphi}^2
\end{eqnarray}
The difference $T^c_{\mu\nu}=<t^\varphi_{\mu\nu}>-T^h_{\mu\nu}$ is due
to the cosmon fluctuations (similar to eq. (\ref{2.4}) and can again be
written in the form
\begin{equation}\label{N6.D}
T^c_{00}=\rho_c\ ,\quad T^c_{ij}=p_c\bar g_{ij}\end{equation}
It has been discussed in \cite{CDM}.

(2) The evolution equation for the background scalar field also
obtains a contribution $q^\varphi$ from backreaction effects \cite{BR}
\cite{CW2}
\begin{equation}\label{N6.E}
\ddot{\bar\varphi}+3H\dot{\bar\varphi}+\frac{\partial V}{\partial
\varphi}(\bar\varphi)=q^\varphi\end{equation}
For cosmon dark matter the ``incoherence force'' $q^\varphi$ has
been discussed in \cite{CDM}. We note that $q^\varphi$ can also
receive a contribution if the cosmon couples to ``standard''
cold dark matter
\cite{CW2}. Such a contribution would not be affected by structure
formation.

(3) The gravitational energy density $\rho_g$ and pressure $p_g$
can be enhanced as compared to cold dark matter. We discuss this
possible effect in a simple model of a collection of cosmon lumps
\cite{CL}. We expect that the most important features could be
present also for more general nonlinear cosmon field configurations
beyond the specific model considered here.

Let us consider a collection of cosmon lumps (some of them could be associated \cite{CL}
to some of the galaxies\footnote{For our own galaxy a large ``cosmon halo'' seems unlikely
in view of the strong distortion of light trajectories \cite{ETCW}.})
which can be described in  comoving coordinates as
\begin{eqnarray}\label{6.1}
\varphi&=&\bar\varphi(t)+\sum_\ell\delta\varphi_\ell(u_\ell)\nonumber\\
g_{00}&=&-\{1+\sum_\ell(B_\ell(u_\ell)-1)\}\nonumber\\
g_{ij}&=&a^2(t)\delta_{ij}\{1+\sum_\ell(C_\ell(u_\ell)-1)\}\nonumber\\
g_{0i}&=&0\end{eqnarray}
Here $\vec x_\ell$ is the comoving coordinate of the center of the lump
$\ell$
\begin{equation}\label{6.2}
u^2_\ell=a^2(\vec x-\vec x_\ell)^2\end{equation}
and $\bar\varphi(t)$ is the cosmological background value of the cosmon field
$\varphi$ which leads to homogenous quintessence.
We assume that the lumps are well separated such that $\delta\varphi_
\ell,
B_\ell$ and $C_\ell$ can be determined from the coupled gravity-scalar field
equation for a single (spherically symmetric) lump. This system has been
discussed in \cite{CL} and we concentrate on the
``halo region'' which may give an important contribution to the energy
momentum tensor. In this region we can
approximate\footnote{Note that the singularities at $u_\ell=R_{H,\ell}/e$
correspond  to pointlike singularities at $r_\ell=0$ in Schwarzschild
coordinates.} $(R^2_{H,\ell}/e^2<u^2_\ell\leq R^2_{H,\ell})$
\begin{eqnarray}\label{6.3}
C_\ell&=&\frac{R_{H,\ell}^2}{u^2_\ell}\ln^2\left(\frac
{eu_\ell}{R_{H,\ell}}\right)
\nonumber\\
B_\ell&=&1+\frac{1}{|\gamma_\ell|}\ln\left(\frac{C_\ell u_\ell^2}{R_{H,\ell}^2}\right)
=1+\frac{1}{|\gamma_\ell|}\ln\left[\ln^2\left(\frac{eu_\ell}{R_{H,\ell}}
\right)\right]\nonumber\\
\delta\varphi_\ell&=&\gamma_\ell M\ln B_\ell\end{eqnarray}
where the scale $R_{H,\ell}$ can be associated with the
radius of the halo and $1/|\gamma_\ell|=v_{rot}^2$ is associated
to the rotation velocity of objects in circular orbits
within the halo. The spherically symmetric solution of the coupled
gravity-scalar system in empty space has indeed two integration constants
$(R_{H,\ell},\gamma_\ell)$. The total mass of the object can be expressed
in terms of these constants \cite{CL}. We consider here small
values of $1/|\gamma_\ell|$ which correspond to realistic rotation
velocities and halo extensions of galaxies \cite{CL}.
While $B_\ell$ is close to one except for the
vicinity of the singularity , we
observe that $C_\ell$ deviates substantially from one within the halo.
As a consequence, $h_{ij}/a^2$ is of the order one within the halo
region and we expect substantial contributions to the backreaction
effects!

For cosmon lumps the spacelike components of the
energy momentum tensor are important. This contrasts with stars.
For a static lump the time derivative of the scalar field vanishes,
and one finds for a single cosmon lump
\begin{eqnarray}\label{6.4}
\rho_\varphi&=&-t^0_0=V(\varphi)+\frac{1}{2}\partial^\mu\varphi\partial_\mu
\varphi=V(\varphi)+\frac{1}{2C}\left(\frac{\partial\varphi}{\partial u}\right)^2\nonumber\\
p_\varphi&=&\frac{1}{3}t^i_i=-\frac{1}{3}\big(\rho_\varphi+2V(\varphi)\big)\end{eqnarray}
These relations are easily generalized to a collection of well
separated lumps. The cosmon part of the energy
momentum tensor obeys the equation of state
\begin{equation}\label{6.5}
p_c=w_c\rho_c\ ,\quad w_c=-\frac{1}{3}(1+2<\Delta V>/\rho_c)\end{equation}
Here we have kept only the contribution from the
inhomogenous fluctuations
and $\Delta V$ is the difference between the local value of the
cosmon potential and the homogenous cosmological value\footnote{For
a single cosmon lump the sign $\Delta V$ may be positive or negative,
depending on the sign of $\partial\varphi/\partial u$.}.
For simplicity we will later concentrate on the case where $<\Delta V>$
can be neglected such that $w_c=-1/3$. Within the halo the potential contribution is indeed small. More generally,
the background potential $V(\bar\varphi)$ is small as compared
to the local energy densities
such that effectively $\Delta V\geq 0$. This implies that static
lumps lead to a negative cosmon equation of state, $w_c\leq -1/3$.

We next turn to the gravitational contribution. For a spherically
symmetric static lump we can again use the relations (\ref{N5.1}) and
(\ref{N5.2}) for the total energy density and pressure. In particular, far
away from the lump the solution approaches the standard Schwarzschild
solution \cite{CL} and we infer that the total integrated pressure
vanishes $(u_c=R_H/e)$
\begin{equation}\label{N6.Q}
\hat P(u\to\infty)=\int dV(p_c+p_g+p_M)=0
\end{equation}
This implies a cancellation between a negative cosmon and positive
gravitational contribution!
(We also have included possible matter $(p_M,\rho_M)$ in the ``bulk''
of the object.) A partial cancellation also happens for the energy
density
\begin{equation}
m(u\to\infty)=\int dV(\rho_c+\rho_g+\rho_M)=m
\end{equation}
Indeed, if $\gamma$ is large, the Schwarzschild radius $R_s=m/(8\pi M^2)\approx 2R_H
/|\gamma|$ is small compared to the halo radius $R_H$ and this is
equivalent to a substantial cancellation between $\rho_c$ and $\rho_g$.

For an understanding of these cancellations in more detail it is instructive
to study the gravitational contribution in the linear approximation. Since
the coordinates used for the metric (\ref{6.1}) are not harmonic,
the formulae (\ref{3.5}) and (\ref{3.8}) receive
corrections.\footnote{Alternatively, one may translate the metric (\ref{6.1})
into harmonic coordinates.} Being only interested in the qualitative
features we neglect these corrections here. This
yields for
the gravitational energy density and pressure
\begin{eqnarray}\label{6.6}
\rho_g&=&<\frac{1}{4}h^{\mu\nu}\delta s_{\mu\nu}+\frac{1}{8}
h\bar g^{\mu\nu}\delta s_{\mu\nu}+h^{\mu 0}\delta s_{\mu 0}+h
\delta s_{00}>\nonumber\\
p_g&=&-<\frac{1}{12}h^{\mu\nu}\delta s_{\mu\nu}+\frac{5}{24}
h\bar g^{\mu\nu}\delta s_{\mu\nu}\nonumber\\
&&+\frac{1}{3}h^{\mu i}\delta s_{\mu i}-\frac{1}{3} h\bar g^{ij}\delta s_{ij}>
\end{eqnarray}
If the potential term can be neglected, the cosmon lumps obey
$\delta s_{\mu 0}=0$ such that
\begin{eqnarray}\label{6.7}
\rho_g&=&\frac{1}{8}<\left(2h^{ij}+h\bar g^{ij}
\right)\delta s_{ij}>,\nonumber\\
p_g&=&\frac{1}{8}<\left(-\frac{10}{3}h^{ij}+h\bar g^{ij}\right)
\delta s_{ij}>\end{eqnarray}

We evaluate the above expression in coordinates adapted to
the present cosmological time with
$a=1, \bar g^{ij}=\delta^{ij}$ and assume first
that only the halo region of the lumps contributes effectively to
$\rho_g, p_g, \rho_c$ and $p_c$. We can therefore evaluate the ratios
$p_g/\rho_g$ an $\rho_g/\rho_c$ for a single cosmon lump. In the coordinate
system (\ref{6.1}) one has (neglecting again $V(\varphi)$)
\begin{equation}\label{6.8}
\delta s_{ij}=\partial_i\varphi\partial_j\varphi=\frac{x_ix_j}{u^2}
\left(\frac{\partial\varphi}{\partial u}\right)^2\end{equation}
With $h_{ij}=(C-1)\delta_{ij}=\frac{1}{3}h\delta_{ij}$ one finds
\begin{equation}\label{6.9}
p_g=-\frac{1}{15}\rho_g\end{equation}
and the gravitational energy density $\rho_g$ reads
\begin{equation}\label{6.10}
\rho_g=\frac{5}{8}<(C(u)-1)\left(\frac{\partial\varphi}{\partial u}\right)^2>
\end{equation}
This is to be compared with the energy density in the cosmon field
\begin{equation}\label{6.11}
\rho_c=\frac{1}{2}<C^{-1}(u)\left(\frac{\partial\varphi}{\partial u}\right)^2
>\ ,\quad p_c=-\frac{1}{3}\rho_c\end{equation}
The average in eqs. (\ref{6.10}) and (\ref{6.11}) is given as
an integration over the volume of the lump
\begin{equation}\label{6.12}
<T(u)>=3\int^{R_H}_{u_b}du u^2T(u)/(\rho^3_H-u_b^3)\end{equation}
where $u_b$ corresponds to the radius of the ``bulk'' of the
galaxy and must be larger than the critical value $u_c=R_H/e$ for the
central singularity.
(We recall here that the averaging needs to be done with respect
to the background metric $\bar g_{\mu\nu}$ such that the volume is
just the cartesian volume in the coordinates $\vec x$.
It does not involve the ``microscopic volume'' which would have
an additional factor $\sqrt g=B^{1/2}C^{3/2}.$)

We note that $C(u)$ becomes smaller than one inside the halo
such that $\rho_g$ is indeed negative and $p_g$ positive. (The numerical prefactors will be altered if eqs. (\ref{3.3}) and (\ref{3.2}) are used
instead of eqs. (\ref{3.5}) and (\ref{3.8}).)
This demonstrates how the cancellation between positive
$\rho_c$ and negative $\rho_g$ becomes visible
already in the linear approximation. With $B\approx 1$ and
\begin{equation}\label{6.13}
\left(\frac{\partial\varphi}{\partial u}\right)^2\approx
\frac{4 M^2}{u^2\ln^2(u/u_c)}\end{equation}
we find that the integrands relevant for $\rho_g$ and $\rho_c$, respectively,
can be characterized by
\begin{eqnarray}\label{6.14}
I_{\rho_g}&=&u^3(C(u)-1)\left(\frac{\partial\varphi}{\partial u}\right)^2
=4M^2\left\{\frac{R_H^2}{u}-\frac{u}{\ln^2
(u/u_c)}\right\}\nonumber\\
I_{\rho_c}&=&u^3C^{-1}(u)\left(\frac{\partial\varphi}{\partial u}\right)^2=
\frac{4M^2u^3}{R_H^2\ln^4(u/u_c)}\end{eqnarray}
They are both dominated by the region $u\to u_b$. We conclude
that the energy density and pressure are actually dominated
by the interior of the halo and/or by the bulk. An assumption
about a halo domination is actually not justified, nor is the
linear approximation for the computation of $\rho_g$ and $p_g$.

Nevertheless, the need of a large cancellation between a positive cosmon
energy denstiy and a negative gravitational energy density remains
true for large $|\gamma|$, irrespective of the shortcomings of the
above calculation. Already the integration of the cosmon energy
density over the halo exceeds the total mass by a large factor
$>R_H/R_s\approx|\gamma|$. The total sum (\ref{N5.2}) can only
be balanced by a negative gravitational energy density of almost
equal (averaged) size! We may summarize our discussion be
extracting the following general features for large cosmon
fluctuations: The cosmon energy density $\rho_c$ is positive
and the pressure $p_c$ negative, typically with $p_c\approx
-\rho_c/3$. This is accompanied by a negative gravitational
energy density $\rho_g$ and positive gravitational pressure $p_g$.
For static isotropic configurations large cancellations occur both
for $\rho_c+\rho_g$ and $p_c+p_g$, implying $p_g\approx-\rho_g/3$.
(This differs from eq. (\ref{6.9}) which involves unjustified
approximations.) For more general, in particular non-static,
large cosmon fluctuations the detailed balance between
gravitational and cosmon contributions may not occur anymore.
It is plausible, however, that the above findings about the
sign of the various contributions remains valid.

A very simple, but perhaps important, observation states
that the cosmon pressure $p_c$ is likely to be negative. Indeed,
for large fluctuations we may neglect the subtraction of
the potential and kinetic energy of the background field $\bar\varphi$.
The cosmon pressure is then given by
\begin{equation}\label{6.15}
p_c=<-V(\varphi)-\frac{1}{6}g^{ij}\partial_i\varphi\partial_j\varphi+\frac{1}{2}\dot\varphi^2>\end{equation}
We observe a negative contribution from the gradient term reflecting spatial
inhomogeneities of $\varphi$. Also the contribution of the potential
is negative and only a fast time variation could cancel these two
negative contributions.

Imagine now a period in the cosmological evolution where the
cosmon fluctuations become substantial and their negative pressure
is not (or only partially) cancelled by the pressure of metric
fluctuations. The cosmological evolution would then be substantially
affected by the negative pressure of cosmon dark matter. Furthermore,
the cancellation between cosmon and gravitational energy density could
be more effective than for the pressure. This could lead to a
situation where the total energy momentum tensor is dominated
by cosmon dark matter and quintessence with a substantially
negative equation of state $w$
\begin{equation}\label{6.16}
w=p/\rho\approx\frac{p_c+p_g+p_h}{\rho_c+\rho_g+\rho_h}\end{equation}
In fact, the pressure of dark energy $p_h$ could also turn negative
if the potential dominates over the kinetic energy during such
an epoch. If $w$ becomes smaller than -1/3, the expansion of the universe
accelerates
\begin{equation}\label{6.17}
\frac{\ddot a}{a}=\dot H+H^2=-\frac{\rho}{12M^2}(1+3w)\end{equation}
It is tempting to speculate that such a situation may occur towards
the end of structure formation. The dominant contribution to $p_c+p_g$
may arise from cosmon inhomogeneities on the scales of clusters or
larger. It is even conceivable that the present acceleration occurs
only effectively for the metric averaged over a volume corresponding
to a redshift $z$ of the order one. At earlier terms it may
have been ``visible'' in the averaged metric relating to a smaller
effective volume.

\section{Conclusion}

As a conclusion, let us turn back to the question asked in the title:
can structure formation influence the cosmological evolution? We have
presented in this paper a few estimates and simple model calculations
within a formalism which describes the backreaction of fluctuations.
We find it unlikely that standard cold dark matter
fluctuations lead to a substantial effect, even though these
fluctuations are today large and strongly nonlinear. The basic
reason is that the averaged Einstein equations are linear in the energy
momentum tensor. The direct contribution of fluctuations in the 
energy density and pressure therefore cancels by virtue of the 
averaging.  An indirect effect of these fluctuations shows up
in the form of induced metric fluctuations. This effect is related
to the gravitational energy density and pressure. We have seen,
however, that the size of this induced fluctuation effect is small
unless a substantial part of the matter is in regions with strong 
and time-varying gravitational fields. Furthermore, we have seen that by a
``cosmic virial theorem'' the gravitational pressure cancels the effect of
the pressure of cold dark matter.

The situation can change drastically in presence of a scalar ``cosmon''
field mediating  quintessence. If the cosmon fluctuations grow
large, their contribution to the backreaction becomes typically
quite sizeable. The averaged Einstein equation, as well as the
averaged scalar evolution equation, are not linear in the cosmon
fluctuations. In contrast to standard cold dark matter large fluctuations
make therefore directly a large contribution to the averaged 
equations. Our computation for cosmon lumps has revealed that
typically the induced gravitational energy density and pressure are
also large. For a collection of static and isotropic cosmon lumps
this ``gravitational backreaction'' cancels the ``cosmon backreaction''
to a high degree. For the pressure one observes a matching of a negative 
cosmon and a positive gravitational contribution. For more general
large cosmon fluctuations, in particular if they are not static, this
cancellation may not be perfect. A large backreaction effect would
then be expected for large cosmon fluctuations. We conclude 
that the backreaction could substantially influence the cosmological
evolution after the time when large cosmon fluctuations have developed.

We have also argued that the equation of state of the combined
cosmon and gravitational fluctuations may be substantially negative.
In this event a growth of fluctuations in the cosmon field
towards the end of structure formation could trigger an acceleration
of the expansion of the universe and provide an answer to the
question why such an acceleration happens ``just now''. Many pieces
of the scenario outlined here are, however, fairly speculative.
In particular, it remains to be seen if a realistic effective
action for the cosmon field can be found such that the cosmon
fluctuations indeed grow large in consistency with present 
observational information.

\end{document}